\documentclass[sn-mathphys,Numbered]{sn-jnl}

\usepackage{graphicx}%
\usepackage{multirow}%
\usepackage{amsmath,amssymb,amsfonts}%
\usepackage{amsthm}%
\usepackage{mathrsfs}%
\usepackage[title]{appendix}%
\usepackage{xcolor}%
\usepackage{textcomp}%
\usepackage{manyfoot}%
\usepackage{booktabs}%
\usepackage{algorithm}%
\usepackage{algorithmicx}%
\usepackage{algpseudocode}%
\usepackage{listings}%

\theoremstyle{thmstyleone}%
\theoremstyle{thmstyletwo}%

\theoremstyle{thmstylethree}%

\begin{document}

\title[Article Title]{Thermoelectric single-photon detection through superconducting tunnel junctions}


\author*[1]{\fnm{Federico} \sur{Paolucci}}\email{federico.paolucci@pi.infn.it}

\author[2,3]{\fnm{Gaia} \sur{Germanese}}

\author[3]{\fnm{Alessandro} \sur{Braggio}}

\author[3]{\fnm{Francesco} \sur{Giazotto}}

\affil[1]{\orgdiv{INFN}, \orgname{Sezione di Pisa}, \orgaddress{\street{Largo B. Pontecorvo 3}, \city{Pisa}, \postcode{56127}, \country{Italy}}}

\affil[2]{\orgdiv{Department of Physics E. Fermi}, \orgname{University of Pisa}, \orgaddress{\street{Largo B. Pontecorvo 3}, \city{Pisa}, \postcode{56127}, \country{Italy}}}

\affil[3]{\orgdiv{NEST}, \orgname{Istituto Nanoscienze-CNR and Scuola Normale Superiore}, \orgaddress{\street{Piazza San Silvestro 12}, \city{Pisa}, \postcode{56127}, \country{Italy}}}


\abstract{Bipolar thermoelectricity in tunnel junctions between superconductors of different energy gap has been recently predicted and experimentally demonstrated. This effect showed thermovoltages up to $\pm150\;\mu$V at milliKelvin temperatures. Thus, superconducting tunnel junctions can be exploited to realize a passive single-photon thermoelectric detector $TED$ operating in the broadband range 15 GHz - 50 PHz. In particular, this detector is expected to show a signal-to-noise ratio of about 15 down to $\nu=50$ GHz and a operating window of more than 4 decades. Therefore, the $TED$ might find applications in quantum computing, telecommunications, optoelectronics, spectroscopy and astro-particle physics.}



\maketitle

\section{Introduction}\label{sec1}
Superconducting detectors represent the state-of-the-art to reveal single-photons of frequency lower than 1 THz for their high efficiency and energy resolution \cite{Semenov2002}. 
Woefully, the noise and overheating introduced by the active readout signal are usually detrimental for their detection performance and limit the number of operating low-temperature devices in an array. 
A possible solution is provided by passive superconducting thermoelectric detectors $TED$s generating electrical signals after the photon absorption. Indeed, hybrid superconductor/ferromagnet $TED$s have been proposed \cite{Hei2018} thanks to the predicted, but still not experimentally realized, thermoelectricity by extrinsic particle-hole (PH) symmetry breaking induced by spin splitting and filtering \cite{Ozaeta2014}. 
Here, we show an alternative route to generate thermoelectric $TE$ signals in simple tunnel junctions made of different superconductors by spontaneous PH symmetry breaking and to exploit these phenomena to implement a single-photon $TED$ operating to frequencies down to a few tens of GHz.

\begin{figure*}[t!]
	\centering 
	\includegraphics[width=0.8\textwidth]{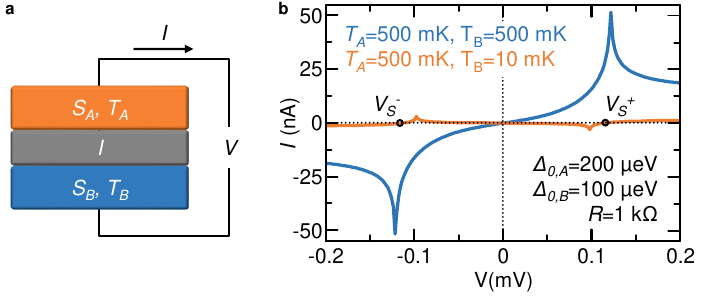}
	\caption{
	\textbf{a} Scheme of a S$_A$IS$_B$ tunnel junction where the voltage drop $V$ and the quasiparticle current $I$ are indicated.
	\textbf{b} $IV$ characteristics of a S$_A$IS$_B$ junction with $\Delta_{0,A}=200\;\mu$eV, $\Delta_{0,B}=100\;\mu$eV, $R=1$ k$\Omega$ calculated for $T_A=T_B=500$ mK (turquoise) and $T_A=500$ mK, $T_B=10$ mK (orange).}
	\label{Fig1}
\end{figure*}

\section{Thermoelectricity in S$_A$IS$_B$ tunnel junctions}\label{sec2}
\subsection{Theory}\label{subsec1}
We study the properties of a S$_A$IS$_B$ tunnel junction with completely suppressed Josephson coupling, where S$_A$ and S$_B$ are the two Bardeen-Cooper-Schrieffer (BCS) superconductors of zero-temperature energy gaps obeying to $\Delta_{0,A}>\Delta_{0,B}$ and I is the insulating barrier [see Fig. \ref{Fig1}(a)]. A smeared PH-symmetric BCS density of states $N_i(\epsilon)=|\Re[(\epsilon+i\Gamma_i)/\sqrt{(\epsilon+i\Gamma_i)^2-\Delta_i(T_i)^2}]|$ (with $i=A,B$ and $\Gamma_i$ the Dynes broadening parameter \cite{Dynes1984}) is assumed for the two leads.
Thus, the quasiparticle current flowing in the systems reads
\begin{equation}
I=\frac{1}{eR}\!\!\int_{-\infty}^{+\infty}\!\!\!\!\!\!\!\!\!\!d\epsilon N_{A}(\epsilon )N_{B}(\epsilon+eV)[f(\epsilon,T_A)-f(\epsilon+eV,T_B)],
\label{eq:iqp}
\end{equation}
where $e$ is the electron charge, $R$ is the junction normal-state resistance, $V$ is the voltage drop, and $f(\epsilon,T_i)=[1+\exp(\epsilon/k_BT_i)]^{-1}$ (with $i=A,B$ and $k_B$ the Boltzmann constant) are the quasiparticle Fermi distributions. 
At the thermal equilibrium ($T_A=T_B$), Eq. \ref{eq:iqp} provides the conventional dissipative behavior, as shown by the turquoise curve in Fig \ref{Fig1}b.
In the presence of a sufficient temperature gradient ($T_A>T_B$), the current flows against the voltage bias, that is the systems shows \textit{absolute negative} conductance (ANC) \cite{Spivak,Ger1}. As a consequence, the S$_A$IS$_B$ tunnel junction is a $TE$ generator \cite{Marchegiani2020,Marchegiani2020_2}. 
In addition, the $IV$ characteristic is \emph{antisymmetric} with $V$ thus entailing the peculiar \emph{bipolar} thermoelectricity, as depicted in Fig \ref{Fig1}b.
Finally, the Seebeck voltages ($V_S^+$ and $V_S^-$) are predicted to be comparable to the value of the matching peak, that is $[\Delta_{A}(T_A)-\Delta_{B}(T_B)]/e\sim |V_{S}^\pm|$ \cite{Marchegiani2020}. 

\begin{figure*}[t!]
	\centering 
	\includegraphics[width=0.8\textwidth]{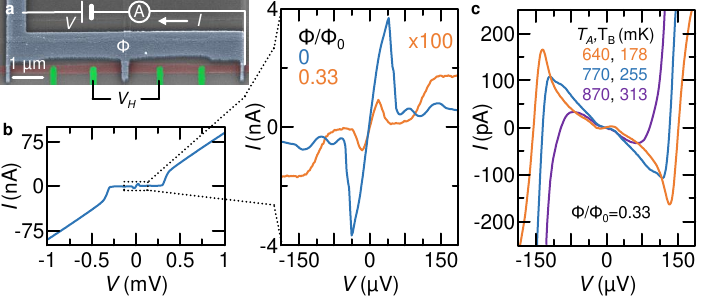}
	\caption{
	\textbf{a} False-color SEM picture of double-loop SQUID ($S_A$ is red and $S_B$ is blue) implementing the $TE$ element and equipped with local tunnel heaters (green). The voltage generator $V$, the current $I$, the heater bias $V_H$ and the magnetic flux $\Phi$ are indicated.
	\textbf{b} $IV$ characteristics of the SQUID measured at $T_A=T_B=30$ mK for $\Phi=0$ (blue) and $\Phi=0.33\Phi_0$ (orange).
    \textbf{c} Thermoelectric $IV$ characteristics recorded at different temperature gradients.}
	\label{Fig2}
\end{figure*}

\subsection{Experiments}\label{subsec2}
Recently, the basic example of a bipolar $TE$ element was realized in the form of a double-loop superconducting quantum interference device (SQUID) \cite{Kemppinen2008,Fornieri2016}, where S$_A$ (red, Al $\Delta_{0,A}=220\;\mu$eV) is coupled to S$_B$ (blue, Al/Cu bilayer $\Delta_{0,B}=80\;\mu$eV) through three insulating aluminum oxide tunnel junctions (see Fig. \ref{Fig2}a). Furthermore, S$_A$ is also equipped with several superconducting tunnel heaters (green, Al $\Delta_{0,H}=200\;\mu$eV) to increase $T_A$.
In this setup, the Josephson coupling can be tuned by the external magnetic flux ($\Phi$), as shown by the 2-probe $IV$ characteristics recorded at $T_A=T_B=30$ mK (see Fig. \ref{Fig2}b).
In particular, at $\Phi=0.33\Phi_0$ the double-loop SQUID permits to suppress the supercurrent down to $0.175\%$ of its maximum value.
Thus, to experimentally demonstrate the predicted $TE$ effect \cite{Marchegiani2020}, the SQUID was flux biased at $\Phi=0.33\Phi_0$ (the Josephson coupling can be practically neglected \cite{Marchegiani2020_3}) at a bath temperature of $30$ mK, while $T_A$ was increased by injecting a power ($P_{H}$) in S$_A$. The power injection causes also the increase of $T_B$, since heat diffuses through the SQUID junctions, but a thermal gradient $\delta T=T_A-T_B$ is established \cite{Germanese2022}. 
Within these conditions, the $IV$ characteristics show ANC for both polarities of $V$ at a \emph{given} temperature profile (see Fig. \ref{Fig2}c) in agreement with the theoretical behavior \cite{Marchegiani2020}.
Differently from conventional thermoelectricity, $|V_{s}|$ monotonically decreases with increasing temperature bias \cite{Germanese2022,Germanese2022_2}.
Finally, the maximum Seebeck coefficient ($\mathcal{S}=V_{S}/\delta T$) is $\sim\pm300\;\mu$V/K, that is near to the maximum theoretically achievable within the exploited architecture \cite{Marchegiani2020,Marchegiani2020_2}.

\begin{figure*}[t!]
	\centering 
	\includegraphics[width=0.8\textwidth]{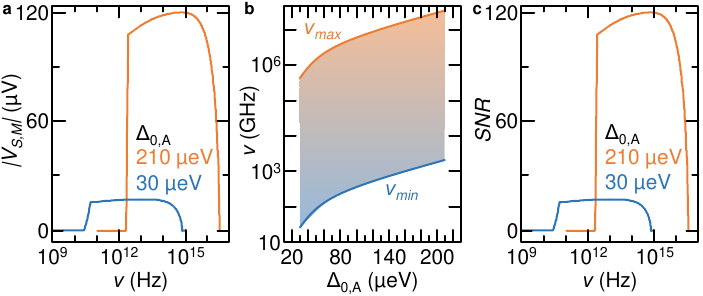}
	\caption{
	\textbf{a} $|V_{S,M}|$ versus $\nu$ calculated for $\Delta_{0,A1}=210\;\mu$eV (orange) and $\Delta_{0,A2}=30\;\mu$eV (blue).
	\textbf{b} Minimum ($\nu_{min}$, blue) and maximum ($\nu_{max}$, orange) operating frequency of the $TED$ versus $\Delta_{0,A}$.
    \textbf{c} $SNR$ versus $\nu$ calculated for $\Delta_{0,A1}=210\;\mu$eV (orange) and $\Delta_{0,A2}=30\;\mu$eV (blue).}
	\label{Fig3}
\end{figure*}

\section{Single-photon superconducting $TED$}\label{sec3}
The single-photon $TED$ \cite{Paolucci2023} exploits the S$_A$IS$_B$ tunnel junction presented in Sec. \ref{sec2} with $\Delta_{0,B}/\Delta_{0,A}=0.5$. Since this system provides two thermovoltages of same absolute value ($|V_S^+|=|V_S^-|$), we will consider only the positive solution $V_S=V_S^+$. 
After absorbing a photon of energy $h\nu$ (with $h$ the Planck constant), S$_A$ reaches a maximum temperature $T_{A,M}$ given by \cite{Moseley1984}
\begin{equation}
    \int_{T_A=T_{bath}}^{T_{A,M}(\nu)} C_{e}(T) \text{d}T = h\nu,
    \label{eq:T}
\end{equation}
where $T_{bath}$ is the bath temperature and $C_e(T)= T \tfrac{\text{d}}{\text{d}T} \mathcal{S}_A (T)$ is the electronic thermal capacity of S$_A$. 
The of S$_A$ is computed as \cite{Rabani2008}
\begin{equation}
\mathcal{S}_A (T)=-2\mathcal{V}_A \mathcal{N}_{F,A} k_B \int_{-\infty}^{\infty} \text{d}\epsilon N_{A}(\epsilon,T)f(\epsilon,T)\ln[f(\epsilon,T)],
\end{equation}
where $\mathcal{V}_A$ and $\mathcal{N}_{F,A}$ are the volume and the density of states at the Fermi energy of S$_A$, respectively. Thus, the open circuit thermovoltage $V_{S,M}$ originated by the photon absorption can be computed by solving Eq. \ref{eq:iqp} for $T_{A,M}(\nu)$ given by Eq. \ref{eq:T} and $T_B=T_{bath}$.
We evaluated the detector output signal by considering two different zero-temperature energy gaps for S$_A$, that is $\Delta_{0,A1}=210\;\mu$eV and $\Delta_{0,A2}=30\;\mu$eV, and a volume $\mathcal{V}_A=2.5\times 10^{-19}$ m$^3$ (see Fig. \ref{Fig3}a). As expected, the $TED$ output signal lowers by decreasing $\Delta_{0,A}$. However, $V_{S,M}$ is almost constant with $\nu$. Contrarily, the detector becomes sensitive to a few tens of GHz single-photons for low gap superconductors. 
The operating frequency window of the $TED$ stays almost constant for different values of the superconducting energy gap, as shown in Fig. \ref{Fig3}b. 
The detector signal-to-noise ratio $SNR$ provides the strength of the output signal relative to the noise. The $SNR$ our $TED$ is given by $SNR[T_{A,M}(\nu)]=V_{S,M}[T_{A,M}(\nu)]/(S_{V}\sqrt{\omega})$, where $\omega=1$ MHz is the measurement bandwidth and $S_{V}=1$ nV$/\sqrt{\text{Hz}}$ is the system voltage noise spectral density. Indeed, we assume noise provided by the voltage pre-amplifier only, since the $TED$ is operated in the idle state in the absence of external signals. 
A $TED$ of gap $\Delta_{0,A2}=30\;\mu$eV is expected to show $SNR\sim15$ for almost the entire operation window (down to $\sim50$ GHz) when operated at $T_{bath}=10$ mK (see Fig. \ref{Fig3}c). Since it is directly linked to $V_{S,M}$, the $SNR$ is directly to the superconducting energy gap of S$_A$.

\section{Conclusions and perspectives}\label{sec4}
In this paper, we envisioned with a concrete proposal the possibility of sizeable thermoelectricity in superconducting tunnel junctions induced by non-equilibrium spontaneous symmetry breaking mechanism \cite{Marchegiani2020,Marchegiani2020_2}. The double-loop SQUID showed bipolar thermovoltages up to $\pm 150\;\mu$V corresponding to a Seebeck coefficient of $\pm 300\;\mu$V/K \cite{Germanese2022,Germanese2022_2}.
This effect was exploited to theoretically implement a passive single-photon $TED$ \cite{Paolucci2023} operating in a wide range of frequencies down to a few tens of GHz. In particular, the $TED$ is predicted to show a $SNR\sim 15$ for 50 GHz single-photons.
To this scope, the $TED$ could be couple to a twins-slot antenna by an impedance matched coplanar waveguide, similarly to $SIS$ detectors that show a relatively high normal-state resistance. \cite{Ezawa2019}
This broadband $TED$ could find practical applications both in quantum technologies, such as telecommunications, optoelectronics and qubits, and in basic science, such as axions search, astronomy and THz spectroscopy.
Finally, bipolar thermoelectricity in superconducting tunnel junctions can find immediate applications in various fields of superconducting quantum technologies, such as a heat engine producing power on a generic load and a persistent $TE$ memory cell written or erased by current injection \cite{Marchegiani2020_3}.

\bmhead{Acknowledgments}
The research leading to these results received partial funding from the EU’s Horizon 2020 research and innovation program under Grant Agreement No. 800923 (SUPERTED), No. 964398 (SUPERGATE) and No. 101057977 (SPECTRUM). A.B. acknowledges PRIN2022 PNRR MUR project NEThEQS (Grant No. 2022B9P8LN) and the Royal Society through the International Exchanges Scheme between the UK and Italy (Grants No. IEC R2 192166.) for partial financial support.

\section*{Declarations}
The authors have no relevant financial or non-financial interests to disclose.

\end{document}